\documentclass[aps,manuscript,preprint,superscriptaddress,]{revtex4-1}
\usepackage[T1]{fontenc}
\usepackage[latin9]{inputenc}
\usepackage{graphicx}

\makeatletter

\@ifundefined{textcolor}{}
{%
 \definecolor{BLACK}{gray}{0}
 \definecolor{WHITE}{gray}{1}
 \definecolor{RED}{rgb}{1,0,0}
 \definecolor{GREEN}{rgb}{0,1,0}
 \definecolor{BLUE}{rgb}{0,0,1}
 \definecolor{CYAN}{cmyk}{1,0,0,0}
 \definecolor{MAGENTA}{cmyk}{0,1,0,0}
 \definecolor{YELLOW}{cmyk}{0,0,1,0}
}

\makeatother

\begin{document}

\title{Neoclassical Pitch-Angle Scattering of Runaway Electrons}

\author{Jian Liu}
\affiliation{School of Nuclear Science and Technology and Department of Modern
Physics, University of Science and Technology of China, Hefei, Anhui
230026, China}
\affiliation{Key Laboratory of Geospace Environment, CAS, Hefei, Anhui 230026,
China}

\author{Yulei Wang}
\affiliation{School of Nuclear Science and Technology and Department of Modern
Physics, University of Science and Technology of China, Hefei, Anhui
230026, China}
\affiliation{Key Laboratory of Geospace Environment, CAS, Hefei, Anhui 230026,
China}

\author{Hong Qin }
\thanks{Corresponding author. hongqin@ustc.edu.cn}
\affiliation{School of Nuclear Science and Technology and Department of Modern
Physics, University of Science and Technology of China, Hefei, Anhui
230026, China}
\affiliation{Plasma Physics Laboratory, Princeton University, Princeton, NJ 08543}
\begin{abstract}
It is discovered that the tokamak field geometry generates a pitch-angle
scattering effect for runaway electrons. This neoclassical pitch-angle
scattering is much stronger than the collisional scattering and invalidates
the gyro-center model for runaway electrons. As a result, the energy
limit of runaway electrons is found to be larger than the prediction
of the gyro-center model and to depend heavily on the background magnetic
field.
\end{abstract}
\maketitle
Runaway electrons in a tokamak are energetic particles accelerated
by the electric field. They cannot be braked by the collisional drag
\cite{Drercer_REorigins_1959}. A large amount of runaway electrons
are produced in tokamaks during fast shutdowns, disruptions \cite{Yoshino_Shutdown_RE,Jaspers_Disruption_RE,Helander_avalanch_2000,Helander_2002,Fulop_magneticThreshold4RE2009,Gill_REref1_2000,Jaspers_RErefs4_1993,Nygren_RErefs5_1997,Parks_RErefs6_1999,Rosenbluth_RErefs7_1997,Yoshino_REandTurbulenceDischarge2000,Tamai_Yoshino_REtermination_JT60U_2002,Lehnen_RMP_PRL_TEXTOR,Finken_RElosses_2007},
or aggressive current drive \cite{Net_Fisch_RevModP1987}. The massive
energy carried by runaway beams poses great danger to plasma-facing
components \cite{Bartels_RE_PFCs1994,Kawamura_PFS_RE1989,Bolt_REref0_1987,Jaspers_REref3_2001}.
Understanding the physics of runaway electrons in toroidal field configurations
is thus critical. The dynamics of runaway electrons involves different
timescales spanning 11 orders of magnitude, which brings difficulties
to both analytical and numerical studies. Gyro-center model is often
applied to tackle the multi-scale problem by averaging out the fast
gyro-motion and has produced fruitful results \cite{Martin_Momentum_RE_1998,Martin_Energylimit_RE_1999,LiuJian_RE_Positron_2014,Bakhtiari_Momentum_RE_Fb_2005,Guan_Qin_Sympletic_RE,Neoclassical_Drift_report,Papp_Fulop_Helander_MPF2011,Rax_RE_resonance1993,Andersson_Helander2001}.
According to the gyro-center model, the magnetic moment of runaway
electron is an adiabatic invariant, and the parallel momentum increases
due to the work by the loop electric field. There is no channel of
momentum transfer from the parallel to the perpendicular direction,
except for the collision with background plasmas. In general, the
collisional effect is rather weak for charged particles with high
velocities \cite{Guan_Qin_Sympletic_RE}. For a typical runaway electron,
the collision time is $\tau_{col}\sim0.5\,\mathrm{s}$, which is much
longer than the gyro-period $T_{ce}$ ($\sim10^{-10}\,\mathrm{s}$)
and the transit period $T_{tr}$ ($\sim10^{-8}\,\mathrm{s}$). When
the collisional effect can be neglected, the perpendicular momentum
will monotonically decrease due to the synchrotron radiation of the
gyro-motion \cite{Guan_Qin_Sympletic_RE}, and parallel momentum will
monotonically increase to its maximum limit until the electric field
acceleration is finally balanced by the radiation dissipation. The
pitch-angle scattering due to collisions will transfer a small amount
of energy from the parallel direction to the perpendicular direction.
This small collisional effect keeps the runaway electrons energetic
in the perpendicular direction, but does not modify the energy limit
in the parallel direction by much. The gyro-center model predicts
that the energy limit of runaway electrons does not depend on collisions
and the magnitude of background magnetic field \cite{Martin_Momentum_RE_1998}.

Contrary to the common wisdom, our analysis shows that the gyro-center
model is not valid for runaway electrons. This is because they move
with the speed of light in the parallel direction, the local magnetic
field they see changes by a large amount during one gyro-period. Therefore,
the basic assumption for the gyro-center model, i.e., the magnetic
field is approximately constant in one gyro-period, breaks down. In
fact, the magnetic moment $\mu$ is no longer an adiabatic invariant.
Similar non-conservation of $\mu$ has been observed in the presence
of magnetic turbulence with wavelength comparable to the gyro-radius
\cite{Dalena_mu_Nonconserve_2012,Dalena_mu_Nonconserve_3Dturbulence}.

\begin{figure}
\includegraphics[scale=0.6]{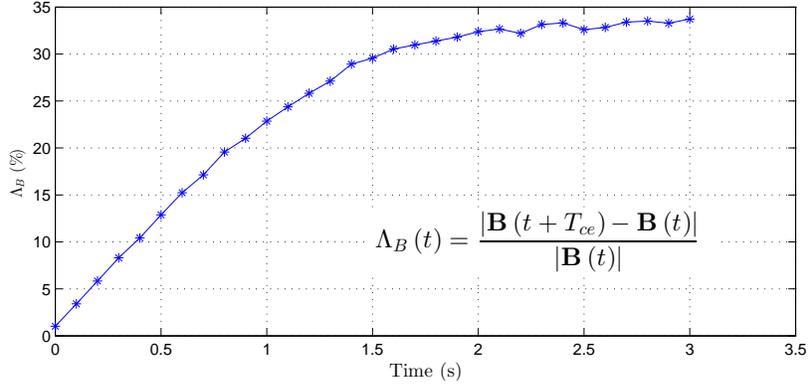}

\protect\caption{The change ratio of background magnetic field $\Delta{}_{B}$ during
one gyro-period at different time of a runaway dynamics.\label{fig:1-B_delta}}
\end{figure}

In this letter, we abandon the gyro-center model and study the multi-timescale
runaway dynamics by numerically solving the dynamical equations of
runaway electrons directly in the six-dimensional phase space. Long-term
simulation results confirm that the gyro-center model is indeed invalid,
see Fig.\,\ref{fig:1-B_delta}. More than one hundred billion time
steps are required in the simulation. To eliminate the coherent accumulation
of numerical errors from each time step, which is usually a show-stopper
for long-term simulations, we utilize the newly developed volume-preserving
algorithm (VPA) for relativistic particles \cite{Ruili_VPA_2015}.
The VPA can guarantee the long-term accuracy by preserving the phase-space
volume of the original physical system. Its long-term conservativeness
and stability have been verified.

Taking the advantage of the VPA method, we discovered that there exists
a new pitch-angle scattering mechanism, which transfers momentum of
runaway electrons between the parallel and perpendicular directions.
It arises from the full orbit dynamics in the toroidal geometry of
a tokamak, hence the name of neoclassical scattering. The neoclassical
pitch-angle scattering process is about a million times faster than
the collisional pitch-angle scattering, resulting in a rapid transfer
of the parallel momentum, gained from the loop electric field, to
the perpendicular direction. As an important result, the simulation
study indicates a new energy limit for runaway electrons, which is
higher than the result from the gyro-center model and varies with
the magnitude of the background magnetic field. This unexpected neoclassical
pitch-angle scattering effect for runaway electrons and its important
consequence are the subjects of this letter.

First, we introduce the physical model. When focusing on the long-term
dynamics of runaway electrons in a tokamak, we take into account the
background magnetic field, the loop electric field, and the electromagnetic
radiation. The dynamical equations of runaway electrons are
\begin{eqnarray}
\frac{\mathrm{\mathrm{d\mathbf{x}}}}{\mathrm{d}t} & = & \mathbf{v},\label{eq:1}\\
\frac{\mathrm{\mathrm{d\mathbf{p}}}}{\mathrm{d}t} & = & -\mathrm{e}\left(\mathbf{E}+\mathbf{v}\times\mathbf{B}\right)+\mathbf{F}_{R},\label{eq:2-1}\\
\mathbf{p} & = & \gamma\mathrm{m}_{0}\mathbf{v},
\end{eqnarray}
where $\mathbf{x}$, $\mathbf{v}$, $\mathbf{p}$ denote the position,
velocity and mechanical momentum of a runaway electron, $\mathrm{e}$
denotes the unit charge, $\mathrm{m}_{0}$ is the rest mass of electron,
$\mathbf{E}$ and $\mathbf{B}$ are the electric and magnetic field,
and the Lorentz factor $\gamma$ is defined as
\begin{equation}
\gamma=\sqrt{1+\frac{p^{2}}{\mathrm{m}_{0}^{2}\mathrm{c}^{2}}}=\frac{1}{\sqrt{1-\left(v/\mathrm{c}\right)^{2}}}\,.\label{eq:2}
\end{equation}
The effective electromagnetic radiation drag force $\mathbf{F}_{R}$
is
\begin{equation}
\mathbf{F}_{R}=-P_{R}\frac{\mathbf{v}}{v^{2}}\,,\label{eq:3}
\end{equation}
where $P_{R}$ is the radiation power determined by \cite{Jackson_electrodynamics}
\begin{equation}
P_{R}=\frac{q_{e}^{2}}{6\pi\mathrm{\mathrm{\epsilon}_{0}c}}\gamma^{2}\left[\left(\frac{\mathbf{a}}{\mathrm{c}}\right)^{2}-\left(\frac{\mathbf{v}}{\mathrm{c}}\times\frac{\mathbf{a}}{\mathrm{c}}\right)^{2}\right]\,.\label{eq:4}
\end{equation}
Here, $\epsilon_{0}$ is the permittivity in vacuum, $\mathrm{c}$
is the speed of light in vacuum, and $\mathbf{a}=\mathrm{d}\mathbf{v}/\mathrm{d}t$
denotes the acceleration.

In order to solve Eqs.\,(\ref{eq:1})-(\ref{eq:2-1}) numerically,
we have to meet the challenge brought by the multi-scale nature of
the problem. Restricted by the minimal timescale, more than $10^{11}$
time-steps are required to simulate a complete runaway dynamics. Traditional
algorithms, such as the 4th order Runge-Kutta method, are not qualified
for this long-term simulation, because the coherent accumulation of
numerical error over many time-steps leads to incorrect long-term
simulation results. To overcome this difficulty, geometric algorithms
which can bound the global numerical error for all time-steps should
be adopted \cite{Qin_VariatianalSymlectic_2008,HeYang_Spliting_2015,Qin_Boris_2013,Jianyuan_Multi_sympectic_2013,Ruili_GC_canonical_2014,Ruili_VPA_2015}.
The newly developed relativistic VPA \cite{Ruili_VPA_2015} with radiation
drag is utilized in the present study to guarantee the long-term numerical
accuracy.

As a model, the background magnetic field and inductive electric field
are set to be
\begin{eqnarray}
\mathbf{B} & = & \frac{B_{0}R_{0}}{R}\mathbf{e}_{\xi}-\frac{B_{0}\sqrt{\left(R-R_{0}\right)^{2}+z^{2}}}{qR}\mathbf{e}_{\theta}\,,\label{eq:5}\\
\mathbf{E} & = & E_{l}\frac{R_{0}}{R}\mathbf{e}_{\xi}\,,\label{eq:6}
\end{eqnarray}
where $R=\sqrt{x^{2}+y^{2}}$, $\xi$, and $z$ are radial distance,
azimuth, and height of the cylindrical coordinate system respectively,
$\mathbf{e}_{\xi}$ and $\mathbf{e}_{\theta}$ are the unit vectors
along toroidal and poloidal directions, and $q$ denotes safety factor.
Without loss of generality, we use the parameters of EAST \cite{Wu_East_2007}.
The major radius is $R_{0}=1.7\,\mathrm{m}$, the safety factor is
$q=2$, the central magnetic field is $B_{0}=3\mathrm{\, T}$, and
the loop electric field is $E_{l}=0.2\,\mathrm{V}/\mathrm{m}$. The
initial parallel and perpendicular momentum of a typical runaway electron
are set to be $p_{\parallel0}=5\,\mathrm{m_{0}c}$ and $p_{\perp0}=1\,\mathrm{m_{0}c}$,
and the initial position is $R=1.8\,\mathrm{m}$ and $\xi=z=0$. The
time-step of simulation is set to $\Delta t=1.9\times10^{-12}s$,
which is about 1\% of the gyro-period.

To verify that the condition for gyro-center approximation is not
satisfied for runaway electrons in tokamaks, we record the change
ratio of the magnetic field $\Delta{}_{B}\left(t\right)$ that one
runaway electron samples during one gyro-period at different time
$t$, see Fig.\,\ref{fig:1-B_delta}. The change ratio is defined
as

\begin{equation}
\Delta{}_{B}\left(t\right)=\frac{\left|\mathbf{B}\left(t+T_{ce}\right)-\mathbf{B}\left(t\right)\right|}{\left|\mathbf{B}\left(t\right)\right|}\,,\label{eq:7}
\end{equation}
where $T_{ce}=2\pi\gamma\mathrm{m}_{0}/eB$ is the gyro-period. For
the gyro-center approximation to be valid, the variation of the magnetic
field a particle samples during one gyro-period should be small. However,
simulation results show that the change ratio $\Delta{}_{B}$ increases
monotonously with the runaway energy. Its value increases to 10\%
after $0.4\,\mathrm{s}$ and exceeds 30\% after $1.7\,\mathrm{s}$,
which can no longer be taken as a small value. This is mainly because
the velocity of runaway electron approaches the speed of light, and
the gyro-period $T_{ce}$ is proportional to the Lorentz factor $\gamma$.
The runaway electron travels a long distance along the toroidal direction
during each gyro-period, which leads to a large value of $\Delta{}_{B}$.

\begin{figure}
\includegraphics[scale=0.6]{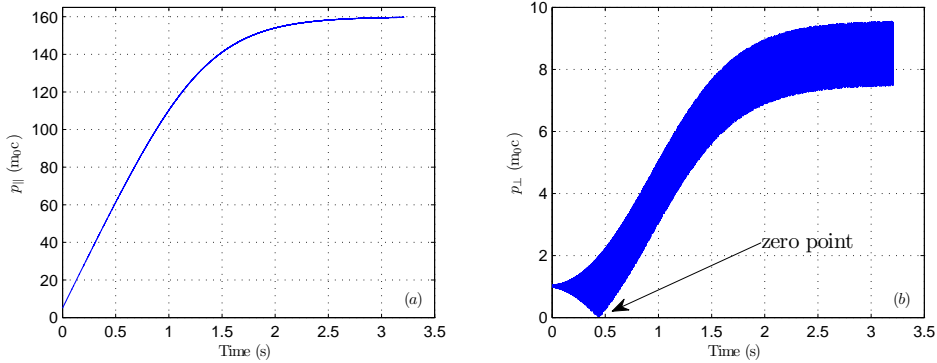}

\protect\caption{The evolution of (a) parallel momentum and (b) perpendicular momentum
of a runaway electron. The initial position of the electron is at
$R=1.8\,\mathrm{m}$, $\xi=z=0$, and the initial momentum is $p_{\parallel}=5\,\mathrm{m_{0}c}$
and $p_{\perp}=\mathrm{m_{0}c}$. The center magnetic field is $B_{0}=3\mathrm{\, T}$
and the loop electric field is $E_{l}=0.2\,\mathrm{V}/\mathrm{m}$.\label{fig:2-Moments}}
\end{figure}

The momentum evolution of runaway electrons is plotted in Fig.\,\ref{fig:2-Moments}.
The increase of the parallel momentum is similar to the results from
the gyro-center model \cite{Guan_Qin_Sympletic_RE,LiuJian_RE_Positron_2014},
but the evolution of the perpendicular momentum is very different.
The perpendicular momentum grows with rapid oscillations, even in
the absence of collisions, until approaching a maximum after about
$2.5\,\mathrm{s}$. Because the parallel momentum is relatively large,
its oscillation is less prominent. The oscillation is an effective
scattering process transferring the parallel momentum to the perpendicular
direction and altering the pitch-angle. As explained later, this effective
pitch-angle scattering roots in the geometric configuration of the
field. It is thus a neoclassical pitch-angle scattering effect. The
evolution of the perpendicular momentum exhibits four stages: (a)
a rapid oscillation is developed initially; (b) then its absolute
value reaches the zero point; (c) it grows quickly after passing the
zero point, and (d) it saturates. At the saturation, though the pitch-angle
still varies quickly due to the neoclassical scattering, the average
of the perpendicular momentum does not change. The timescale of neoclassical
pitch-angle scattering is $5\times10^{-7}\,\mathrm{s}$, which is
about $10^{6}$ times faster than the collisional scattering, and
much more momentum can be transferred to the perpendicular direction
through the neoclassical scattering than the collisional effect as
calculated in Ref \cite{Guan_Qin_Sympletic_RE}.

\begin{figure}
\includegraphics[scale=0.7]{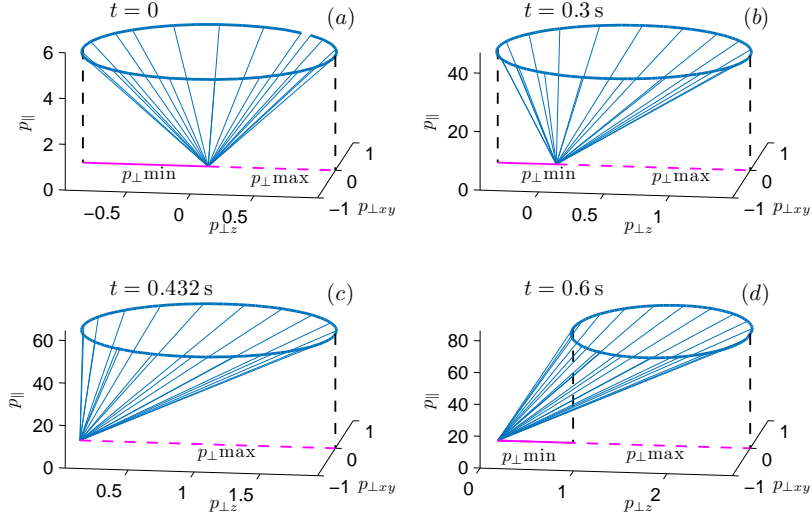}

\protect\caption{Motion of the momentum vector during one gyro-period at different
time. The maximal and minimal perpendicular momenta, $p_{\perp max}$
and $p_{\perp min}$, during each gyro-period are marked by dashed
line. \label{fig:4-Snapshots-of-momentum}}
\end{figure}

The seemingly complex evolution curves of the perpendicular momentum
are dominated by the neoclassical scattering, which can be analyzed
by looking at the variation of the momentum vector. We choose the
parallel moment $p_{\parallel}$, the z-component of the perpendicular
momentum $p_{\perp z}$, and the projection of perpendicular moment
in the $x-y$ plane $p_{\perp xy}$ as the three coordinates for the
moment. They satisfy $p_{\perp xy}^{2}+p_{\perp z}^{2}+p_{\parallel}^{2}=p^{2}$.
In Fig.\,\ref{fig:4-Snapshots-of-momentum}, snapshots of the momentum
vector within one gyro-period at different moments are plotted in
the momentum space. During each gyro-period, the tip of the momentum
vector moves approximately along a circular orbit. The minimal and
maximal value of $p_{\perp z}$ are marked within each gyro-period.
The circular orbit is first elongated while the center of circle shifts
to $p_{\perp z}$ direction in the perpendicular plane. The elongation
of the orbit corresponds to the growth of the oscillation amplitude.
At $t=0.432\,\mathrm{s}$, the orbit touches the $p_{\perp z}=0$
plane, and the zero point of perpendicular momentum appears. Afterwards,
the elongated orbit keeps shifting towards the larger $p_{\perp z}$
direction until approaching a steady state. It is evident that the
variation of $p_{\perp}$ is mainly due to the $z$-component $p_{\perp z}$.

\begin{figure}
\includegraphics[scale=0.47]{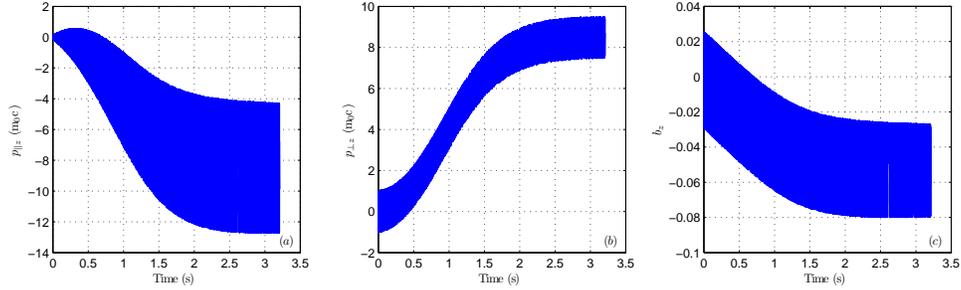}

\protect\caption{Evolution of (a) $z$-component of parallel momentum $p_{\parallel z}$,
(b) $z$-component of perpendicular momentum $p_{\perp z}$, and (c)
$z$-component of unit vector along the magnetic field $b_{z}$.\label{fig:Perp_z}}
\end{figure}

Figure \ref{fig:Perp_z} shows the evolution of the $z$-component
of the parallel momentum $p_{\parallel z}$, the $z$-component of
the perpendicular momentum $p_{\perp z}$, and the $z$-component
of the unit vector along the magnetic field $b_{z}=\mathbf{b\cdot z}=\mathbf{B}\cdot\mathbf{z}/B$.
We find that $p_{\perp z}$ increases, oscillates, and saturates around
$9\, m_{0}c$. The oscillation amplitude of $p_{\parallel z}$ increases
with the absolute value of $p_{\parallel}$. Meanwhile, the evolution
of $p_{\parallel z}$ has the same trend as $b_{z}$, which indicates
that the neoclassical scattering is closely related to the direction
of the local magnetic field. Due to the neoclassical drift \cite{Neoclassical_Drift_report,Guan_Qin_Sympletic_RE},
the transit orbits of runaway electrons drift outwards from the magnetic
axis. As a result, runaway electrons spend more and more time in the
$R>R_{0}$ region. The magnetic field runaway electrons sample then
tilts more towards the negative z direction on average, because of
the helical configuration of the magnetic field lines. Since the parallel
momentum is defined to be $p_{\parallel}=\mathbf{p\cdot b}$, the
change of $\mathbf{b}$ results in a change of $p_{\parallel}$. In
the $z$-direction, the time average of momentum vanishes approximately,
i.e., $\left\langle p_{z}\right\rangle =\left\langle p_{\parallel z}\right\rangle +\left\langle p_{\perp z}\right\rangle =0$.
Therefore, decrease of $\left\langle p_{\parallel z}\right\rangle $
corresponds to an increase of $\left\langle p_{\perp z}\right\rangle $.

\begin{figure}
\includegraphics[scale=0.7]{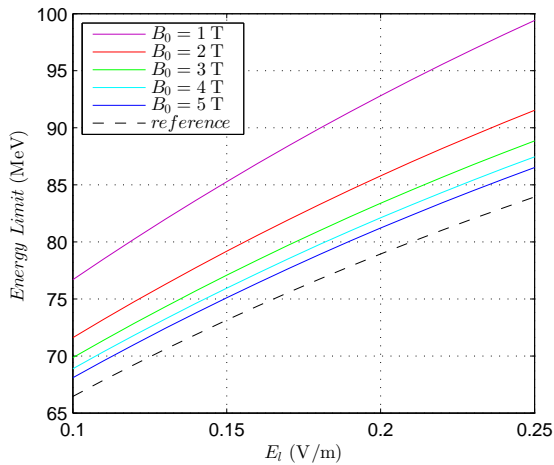}

\protect\caption{Plots of energy limit versus loop electric field $E_{l}$. The black
dashed curve corresponds to the energy limit curve predicted by gyro-center
model \cite{Martin_Momentum_RE_1998}, where collisional effect is
ignorable. The solid curves correspond to energy limits with different
magnetic field intensities. The loop electric field is set to be radially
uniform in order to compare with the gyro-center model.\label{fig:5-Energy-limit}}

\end{figure}

The energy limit is reached at the saturation. The variation of the
energy limit against the loop electric field with different magnetic
field intensities are plotted in Fig.\,\ref{fig:5-Energy-limit},
which clearly shows that the energy limit depends on the intensity
of the background magnetic field. This is different from the conclusion
by Martín-Solís et al. using the gyro-center model \cite{Martin_Momentum_RE_1998}.
For the case of $B_{0}=1\,\mathrm{T}$, our energy limit is about
$20\%$ higher, because more energy is transferred to the perpendicular
direction through the neoclassical pitch-angle scattering. If the
magnetic field is extremely strong, the gyro-center approximation
model will be valid, and our energy limit curve will recover the gyro-center
result.

In summary, long-term simulations using the newly developed volume
preserving algorithm revealed that the full orbit dynamics of a runaway
electron in the tokamak field geometry generates a pitch-angle scattering
effect for runaway electrons. In a typical tokamak, this neoclassical
pitch-angle scattering is about 1 million times stronger than the
collisional scattering and invalidates the gyro-center model for runaway
electrons. As a consequence, the energy limit of runaway electrons
is found to be larger than the prediction by a gyro-center model.
In addition, the theoretical model developed in the present study
shows that the runaway energy limit also depend heavily on the background
magnetic field.
\begin{acknowledgments}
This research is supported by National Magnetic Confinement Fusion
Energy Research Project (2015GB111003, 2014GB124005), National Natural
Science Foundation of China (NSFC-11575185, 11575186, 11305171), JSPS-NRF-NSFC
A3 Foresight Program (NSFC-11261140328), the CAS Program for Interdisciplinary
Collaboration Team, and the GeoAlgorithmic Plasma Simulator (GAPS)
Project.
\end{acknowledgments}

\bibliographystyle{apsrev}

\end{document}